# Message interaction dynamics covary with brain volume across mammalian connectomes

Yan Hao[1], Hannah Lax[2], Marc-Thorsten Hütt,[3] and Daniel J. Graham[4]

[1] Department of Mathematics, Hobart and William Smith Colleges, Geneva, New York, USA
[2] U.S. Department of Veterans Affairs, Boston, Massachusetts, USA
[3] Department of Life Sciences and Chemistry, Constructor University, Bremen, Germany
[4] Department of Psychological Science, Hobart and William Smith Colleges, Geneva, New York, USA

## ABSTRACT

Brain network communication models typically assume that signals propagate independently, despite the high network density and small diameter of mammalian connectomes, where interactions among simultaneously propagating messages are likely. We investigate these interactions using the copy–spread–annihilate (CSA) model, a synchronous Markovian message-passing process in which messages spread through the binarized network while undergoing collision-dependent deletion. Simulations on a large comparative dataset of mammalian connectomes show that CSA dynamics produce robust positively-skewed lognormal distributions of message survival across species. Despite using only binary network topology without spatial embedding, message survival accounts for over half of the variance in brain volume across mammal species and outperforms a broad range of established graph-theoretic measures and an alternative communication model with interacting messages. These relationships persisted when measured paths could not revisit nodes. Degree-preserving switch randomization weakens the relationship between message survival and brain volume, indicating that higher-order structural organization contributes to the observed scaling law. We describe a mechanistic explanation of model dynamics that suggests message creation differences among networks shapes message survival. Together, these findings suggest that interactions among propagating messages constitute an important and underexplored determinant of communication dynamics in mammalian brain networks.

## INTRODUCTION

In contrast to our increasing understanding of representation in the brain, the principles governing intercommunication across the brain's structural network remain poorly understood. This theoretical lacuna is due in part to the fact that mapping a network and following signals across it in the same experiment remains difficult, despite recent advances (e.g., Dorkenwald et al., 2024; MICrONS Consortium, 2025). Nevertheless, we possess enough suggestive evidence about brain networks to address general questions about communication strategies on the connectome.

Existing communication models have largely assumed that propagating messages do not interact with one another. This assumption is surprising because connectomes are densely interconnected, with characteristic path lengths of only a few synaptic traversals (Knoblauch et al., 2016; Lin et al., 2024). Under such conditions, simultaneous messages should encounter one another frequently, making interactions a potentially important determinant of network communication (Graham, 2014; 2021; 2023; Hao and Graham, 2020; Luppi et al., 2024; Hao et al., 2026).

Historically, connectomic studies have investigated how variations in structure could underlie brain function in health and disease (see e.g., Kaiser, 2020; Fornito et al., 2016; Sporns, 2016, for overviews). However, moving from static structural analysis to network dynamics, most existing models assume that the flow of signals over the network is monosynaptic, strongly dependent on edge weight, and free of message interactions. These assumptions are so widespread that they often go unspoken, as in studies of "functional connectivity" (see, e.g., Chiêm et al., 2021).

However, increasing numbers of brain network communication models have sought to capture polysynaptic (multi-hop) exchange of information between and among brain regions along white matter connections (e.g., Fornito et al., 2016; Avena-Koenigsberger et al., 2019; Seguin et al., 2020; Graham et al., 2020; Seguin et al., 2023). These models have mostly examined network structure statistics such as shortest path lengths, path redundancy, and efficiency for passing messages over space, including important work by Avena-Koenigsberger et al. (2019); Vázquez-Rodríguez et al. (2019); Suárez et al. (2020) and Seguin et al. (2018; 2019; 2020). Recent work has also examined information transfer (e.g., Rosas et al., 2019). These studies have helped to elucidate how polysynaptic messaging can be shaped by the complex interplay of network topology, edge weights, and physical connection distances. In addition, some models have examined message interactions; these include competitive cascade models (Misic et al., 2015; Betzel et al., 2026), discussed below, as well as *queueing models* (Misic et al, 2014a,b; Fukushima and Leibnitz, 2025), discussed in the Supplementary Materials.

Here we describe experiments aimed at further elucidating the role of message interactions on brain networks and their potential for shaping message passing strategies and evolutionary scaling of brains across mammal species.

We study the *copy-spread-annihilate* (CSA) model (Hao and Graham, 2020; Hao et al., 2026), a discrete Markovian process which combines spreading of duplicated messages with destructive collisions of messages simultaneously arriving at nodes.

The first component of the system's dynamics models message broadcasting: messages arriving singly at a node replicate themselves to all nearest neighbors. Broadcasting processes, also termed spreading or cascade processes, are emerging as a fundamental component of candidate models of whole-brain function (Betzel et al., 2026; Fakhar et al., 2025, Misic et al., 2015) and have long been a foundational idea in global workspace models (Baars, 1997; Paz, 2022).

The second component involves message deletion. Biological mechanisms for deletion of simultaneously arriving messages could include classical inhibition, as well as XOR-type interactions (Gidon et al., 2020; 2022) and other gating mechanisms (see Olshausen et al., 1993; Steriade and Paré, 2007; Gollisch and Meister, 2010; see Graham, 2023).

Some existing models have examined "competitive" cascades, i.e., situations where colliding messages originating from different injections compete for the right to occupy a given receiver. In Betzel et al. (2026), the message source with the majority of arriving copies "wins" (ties are resolved by a coin toss), while the competitive cascade model of Misic et al. (2015) adds a threshold value that all message delivery cohorts must surpass in order to compete.

While these dynamics do result in limiting message spread, the case of full mutual deletion has not been explored. One consequence of destructive collisions is the limitation of message survival time on the network. This would appear to be problematic for network communication. However, message deletion is a basic feature of engineered communication systems, typically indicating successful message delivery at a target. The crucial requirement in any communication system is for messages to be halted and deleted at the right place at the right time, a fact which has not been operationalized in most existing models. In the brain, this requirement is almost certainly met in the absence of the kinds of addressing and pathfinding mechanisms employed in engineered systems.

Existing cascade models have also examined only single injections ("seeding") at a one or a handful of input nodes at the beginning of the simulation. Such dynamics cannot account for interactions of "old" and "new" messages; such interactions are relatively probable in networks like the brain that possess short characteristic path lengths. For example, competitive cascades persist over a dozen or more discrete timesteps on the fly connectome in the model of Betzel et al. (2026), while the characteristic path length of the network is about five hops. As a first step to understanding these kinds of interactions, the CSA model is driven by new injections each timestep.

*Exponential Distance Rule*

In both the mouse and macaque monkey, tracer-based studies have shown that the distribution of white matter axon lengths follows a decaying exponential distribution (Markov et al., 2011; Ercsey-Ravasz et al., 2013; Horvát et al., 2016). Across pairs of regions the distribution of interareal distances is lognormally distributed and positively skewed. Thus, situating the exponential distribution of wires in a physical brain with lognormally-distributed interareal distances yields a heavy-tailed lognormal-like distribution of tract lengths between areas, an observation known as the *exponential distance rule* (EDR). EDR-like relationships have been observed in imaging-based connectomes in 125 mammal species (Faskowitz et al., 2023); at the single cell level within regions of mammal cortical gray matter (Horvat et al., 2016); and in the *Drosophila* projectome (Jozsa et al., 2024). In *C. elegans*, Kaiser et al. (2009) found a lognormal-like (gamma) distribution of connections lengths based on a model of axon outgrowth, while Piazza et al. (2025) derived lognormal behavior in several connectomes based on an axonal branching model.

The EDR implies that most pairs of brain areas are separated by relatively short tracts, but the distribution has a heavy tail. As with EDR distance relationships, nearby areas are likely to be connected by stronger links (i.e., more fibers), and strength also falls off with a heavy tail, over which connection strength remains fairly constant (Markov et al., 2011).

The EDR has been invoked to suggest that graph topology alone cannot explain brain organization, either in terms of the brain's communication strategies or its evolution (Wang and Kennedy, 2016). As Horvát et al. (2016) argue, the EDR "is a more basic and general property than the description of cortical connectivity as a network at some coarse-grained (e.g., mesoscale) level…[it] constrains the topological structures that connectomes can form across different levels, ranging from the single neuron to the areal level."

However, EDR-like relationships have so far not been investigated in the context of dynamic message interactions. This fact is surprising given that the EDR implies a high potential for signal interaction. Detailed mechanistic models of brain network communication that invoke or replicate EDR have generally used static network statistics and, in the case of spatial embedding, they have used generative models that recapitulate constraints like the EDR. However, these kinds of models cannot capture the emergent, nonlinear effects of dynamic message interactions. It is not clear that "wire length" should be the only or the most important constraint on message-passing and routing in brain networks. Neuronal signals have tunable delays (e.g., via variations in axon diameter; Innocenti et al., 2016) that could offset constraints imposed by wire length. Considered from the point of view of polysynaptic communication, network topology would appear to be a more important consideration than connection distance or strength as far as message interactions are concerned. This is because nodes are presumably where routing occurs, at least at the mesoscale (axonal branching may play a role as well; see Waxman, 1972; Scott, 1977; Winnubst et al., 2019; Linden, 2022).

Here we seek to understand the extent to which brain network topology by itself constrains message passing in the presence of message interactions, without inclusion of spatial embeddings or edge weights. Because EDR influences network topology, we hypothesized that a well-chosen model of message interactions should give rise to EDR-like behavior "for free." This is because firing rates, synaptic weights, synchronous firing, numbers of synaptic contacts, and dendritic bouton sizes follow similar lognormal-like distributions, leading Buzsáki and Mizuseki (2014) to conclude that "skewed (typically lognormal) distributions are fundamental to structural and functional brain organization."

We apply the CSA model to a large set of mammalian connectomes to show that the model produces results in line with past studies of static network structure. Then, we present novel results regarding effects of evolutionary scaling of mammal brains. We investigate CSA model relationships with brain size and compare these to alternative models. We also perform degree-preserving switch randomization to gauge the relative importance of structural regularities and degree distributions in our results. Finally, we provide a mechanistic description based on structure-dynamics interactions to explain our results.

## METHODS

### *Copy-Spread-Annihilate (CSA) Model*

Network dynamics are modeled as a Markovian agent-based process of collisions and spreading activation on empirical connectomes, where nodes serve as synchronous-time message passing agents (see Figure 1). Under this model, a message that meets one or more messages at a node on a given timestep results in all colliding messages being deleted from the system. If a message arrives singly at a node, it survives to the next time step, at which point it passes copies of itself to all the node's nearest neighbors. Each time step, new messages are injected into the system at a fixed number of randomly-chosen unoccupied nodes. The quantity of message injections is termed the *injection rate*. See Figure 1. Message injections represent intrinsic activity as well as external inputs. In past work, we have shown that the CSA model maintains sparse activity on mouse and monkey tracer-based connectomes as injection rate increases (Hao and Graham, 2020[1]). Collisions occur only at nodes rather than along edges meaning that potential ephaptic interactions are not modeled (see e.g., Sheheitli and Jirsa, 2020).

For each injected message, we track all descendant copies throughout the simulation. Message lifetime is defined as the survival time of the longest-lived descendant, following Hao et al. (2026), i.e., the number of timesteps from injection through deletion (inclusive), yielding positive integers with a minimum value of one. This quantity measures the maximum persistence of information generated by a single injection. Hao et al. (2026) showed that the mammal brain network possesses structural regularities that are conducive to promoting message lifetime under CSA dynamics (see Discussion). Here we expand our application of the CSA model in a systematic examination of brain networks across the mammalian class.

[1] The CSA model differs from the "information spreading" model of Hao and Graham (2020) in one subtle respect, namely that the injection rule is slightly different. Here messages can only be injected at open nodes, while in Hao and Graham (2020) new messages (there termed "load") could be injected at any node, including occupied nodes, leading to a collision at injection in that case. However, this difference does not significantly alter the dynamics of the model.

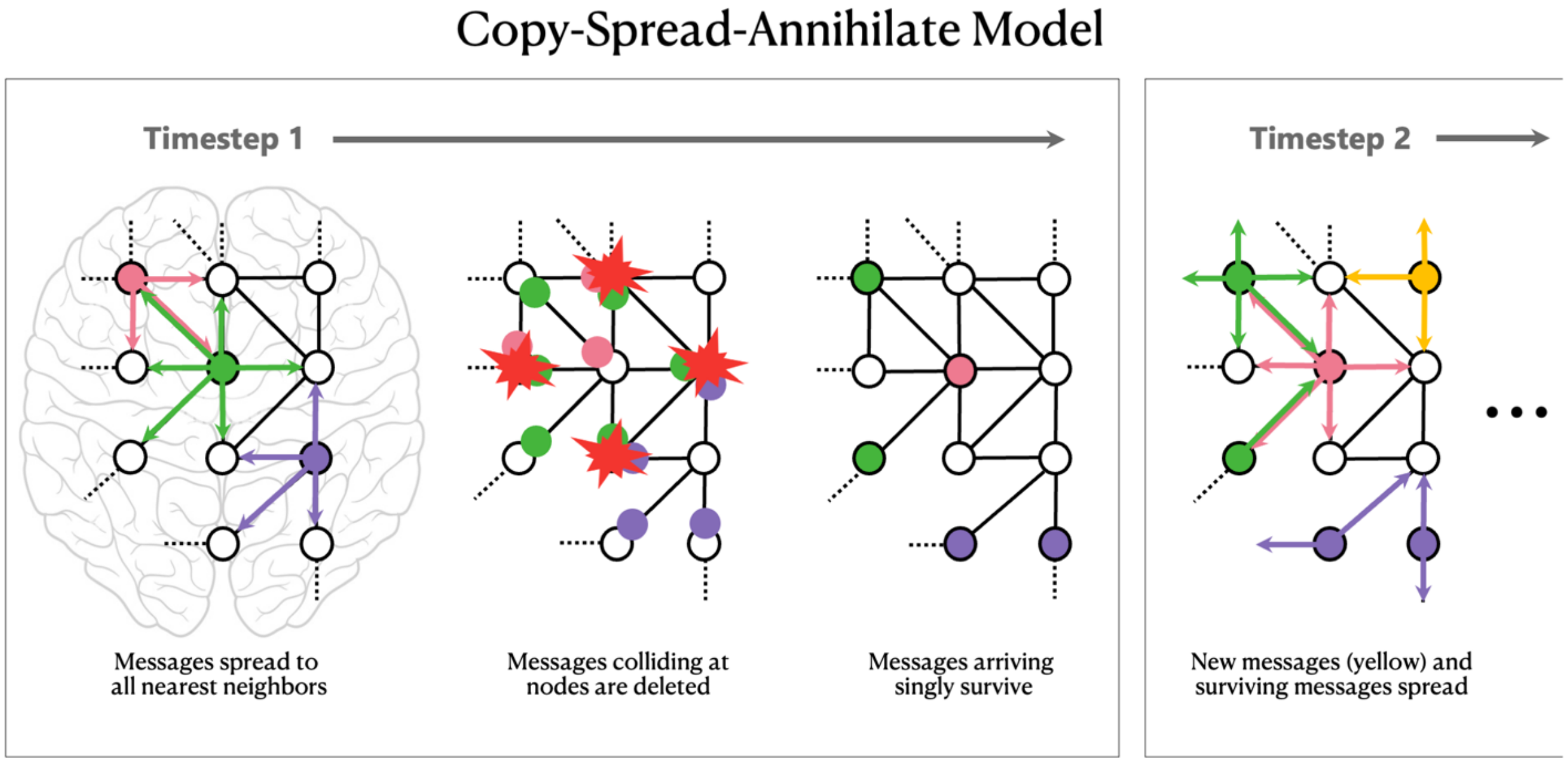

**Figure 1. Copy-spread-annihilate model on the mammalian connectome. On a given timestep, messages in the system spread to all nearest network neighbors. If a message meets one or more other messages at a node, all are deleted. Messages that do not encounter another message survive to the next time step. On the next timestep, a fixed number of open nodes (given by the *injection rate*) are randomly selected to receive a new message (yellow). Then, all messages spread and collide as in timestep 1.**

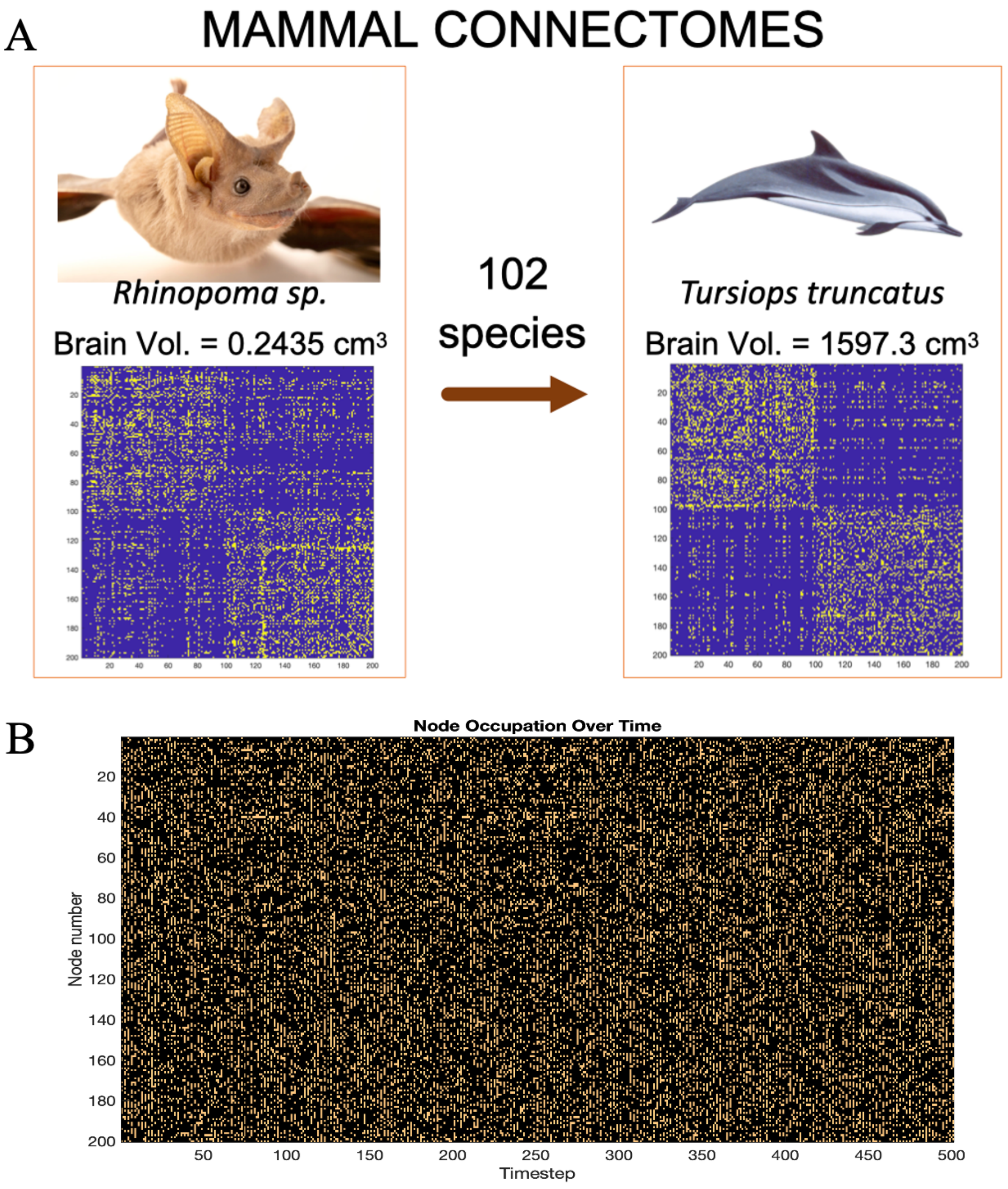

**Figure 2. A. Connectomes used in this study span more than three orders of magnitude in size, from bat to dolphin, comprising 102 different species. Adjacency matrices are shown. B. Visualized simulation shows the position of surviving messages on each timestep across nodes.**

*Connectome Dataset*

We analyzed the MaMI connectome dataset (Puxeddu et al., 2024), comprising approximately 200 diffusion MRI-derived mammalian connectomes representing 102 species and spanning more than three orders of magnitude in brain volume. Connectomes were reconstructed from the diffusion MRI data of Assaf et al. (2020) using the preprocessing pipeline described therein. The Puxeddu et al. (2024) dataset was created using a probabilistic model of connectivity assessed with the network portrait divergence measure (Bagrow et al., 2008; Bagrow and Bollt, 2019) to generate networks with a fixed number of nodes (100 per hemisphere), which lack functional assignment. To eliminate effects of network density, we analyzed the density-0.15 thresholded networks supplied by Puxeddu et al. (2024), which are the highest density available. Five disconnected graphs were excluded, leaving 185 fully connected connectomes. See Figure 2.

*Simulation Workflow*

Each connectome was binarized, giving undirected graphs with equal edge weights of 1. Simulations were run for 2,000 timesteps and repeated 10 times per connectome.

We used an injection rate of 6.5%, i.e., 13 randomly selected unoccupied nodes received new messages; this value was lower than the number of unoccupied nodes on all timesteps of all simulations. The choice of low values of injection rate reflects estimates of typical activity fractions for the mammal brain (Levy & Baxter, 1996; Attwell & Laughlin, 2001; Lennie, 2003; see Hao and Graham, 2020). Injection rates ranging from 5% to 40% produced similar results in all experiments. Higher values lead to almost all messages being deleted after one timestep.

Message statistics were averaged across runs. Results were averaged within the 102 species represented in the dataset and related to brain volume. The same was done for alternative graph-theoretic and dynamic measures. All results presented here are essentially identical when connectomes are not averaged within species. Continuous candidate distributions were discretized to integer bins for likelihood evaluation, while discrete distributions were parameterized on the same support.

Note that no explicit brain-volume, spatial-coordinate, or edge-weight variable is supplied to the CSA simulation.

## RESULTS

<u>Message Survival Distributions</u>

We first asked whether CSA dynamics generate message survival distributions consistent with established large-scale regularities of mammalian brain organization, particularly the heavy-tailed lognormal-like distributions associated with the exponential distance rule.

Throughout this experiment we express message lifetime as *message survival* (measured in edge traversals) to facilitate comparison with studies of interareal distances. It differs in two respects from message lifetime as defined by Hao et al. (2026): first, it implies a network distance traveled, rather than a time period. Second, message survival is not weighted by the number of messages that survive the maximum number of timesteps from injection. In all other respects, it is identical to message lifetime, and all results are very similar when survival is weighted by the number of messages surviving the maximal time from injection in each connectome. When considered as distances, it should be noted that messages can "retrace their steps" and/or travel in loops, meaning that they are not necessarily traveling on geodesic paths between injection and final deletion. However, to better understand relationships of message travel with distance, we also examined the longest loop-free injection-to-collision trajectory from each injection, along with its spatial extent. We also measured message coverage over each graph.

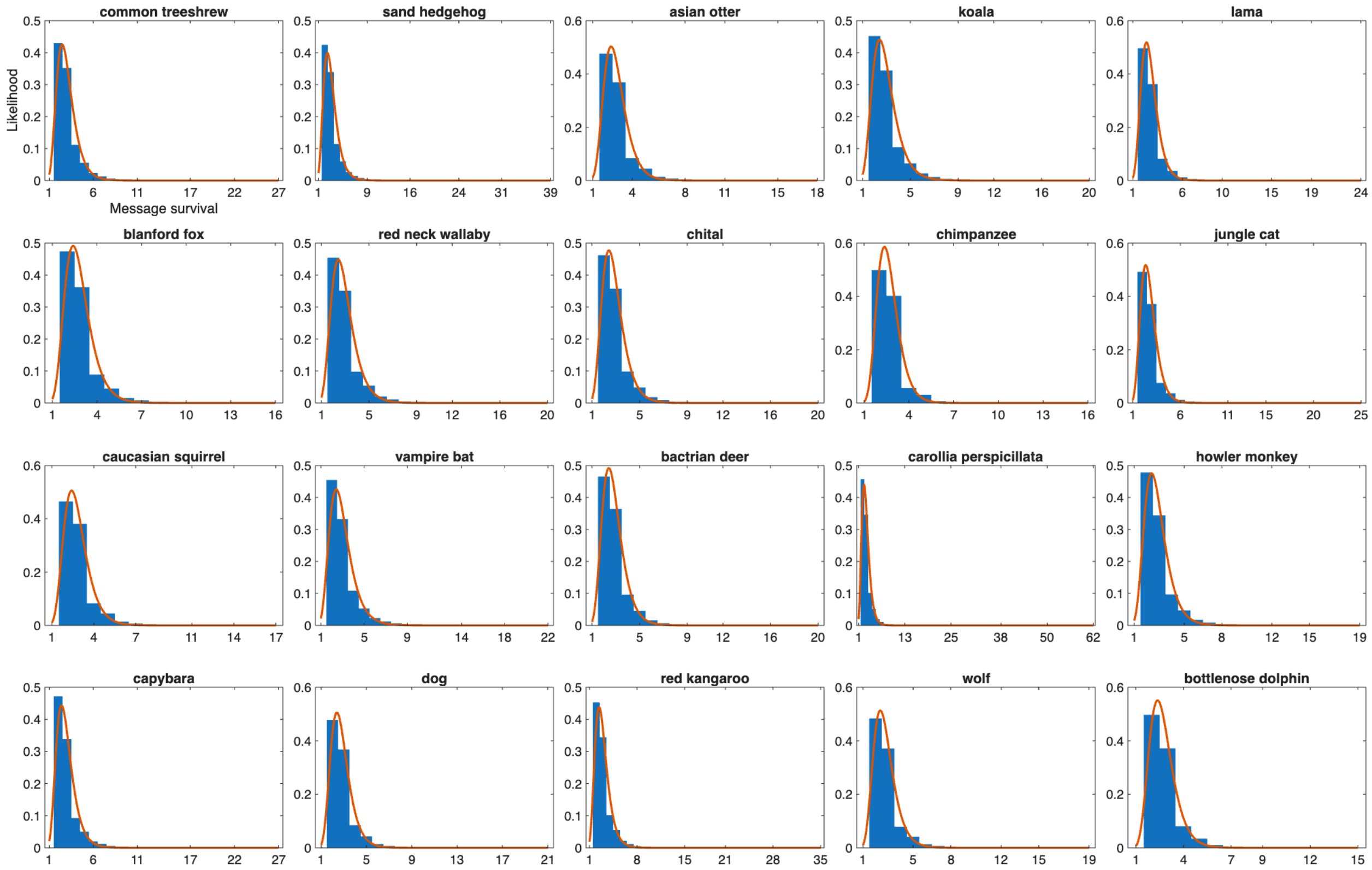

**FIGURE 3. Message survival distributions under the CSA model and lognormal fits for 20 representative species averaged across individuals of that species.**

We fit each distribution with six candidate distributions: lognormal, negative binomial, exponential, geometric, Weibull and gamma. Across 102 mammal species, message survival distributions were overwhelmingly best described by a lognormal distribution. Compared to other candidate distributions, the lognormal model was selected for 99 of 102 species (97.1%) by both AIC and BIC, and for 100 of 102 species (98.0%) by held-out log likelihood. In the remaining species, the only competing model selected was the negative binomial; no species were better fit by exponential, geometric, Weibull, or gamma distributions.

*Colliding Random Walk Model*

As a comparison, we evaluated a colliding random walk (colliding-RW) model in which messages execute unbiased random walks without message copying while obeying the same collision rule and injection process as the CSA model. The colliding-RW model was studied by Hao and Graham (2020) on mammalian connectomes and found to have high and increasing levels of node activity as a function of message injection rate, and low and decreasing sparseness of occupation as a function of injection rate, unlike the CSA model.

The same simulation parameters employed in the CSA model were used for the colliding-RW model. Because there are no message copies made after injection, the colliding-RW survival is simply the average survival tallied over all messages. This measure can be naturally compared to message survival in the CSA model, which likewise measures survival from injection to final collision. However, the colliding-RW model serves as an example at the other end of the redundancy spectrum, rather than as a control model, since messages are produced at much lower rates.

In contrast to CSA, colliding-RW message survival distributions followed exponential distributions within individual species ($R^2 > 0.99$). When pooled across species, mean message survival under the colliding-RW model produced a strongly negatively skewed distribution (skewness = −0.63), a unlike the CSA model, which exhibited a pooled mean message survival distribution with positive skew (skewness = 0.35).

Our results thus provide a picture of the effects of simple message interactions on survival distributions over a large sample of mammal species.

Covariation with Brain Volume

Here we test whether communication dynamics under the CSA model vary systematically with evolutionary size scaling across mammal brains. We also compare our results to standard network statistics measures and the colliding-RW model.

Past analyses of static network structure in thresholded MaMI brains that included spatial embedding, but in the absence of message interactions, found that "as the brain size increases, communication becomes progressively more efficient for regions in closer spatial proximity" (Puxeddu et al., 2024). Likewise, Hilgetag et al. (2016) investigated similarities in laminar organization in the primate cortex, which they showed generates connection patterns on the Felleman-Van Essen connectome that are predominantly local.

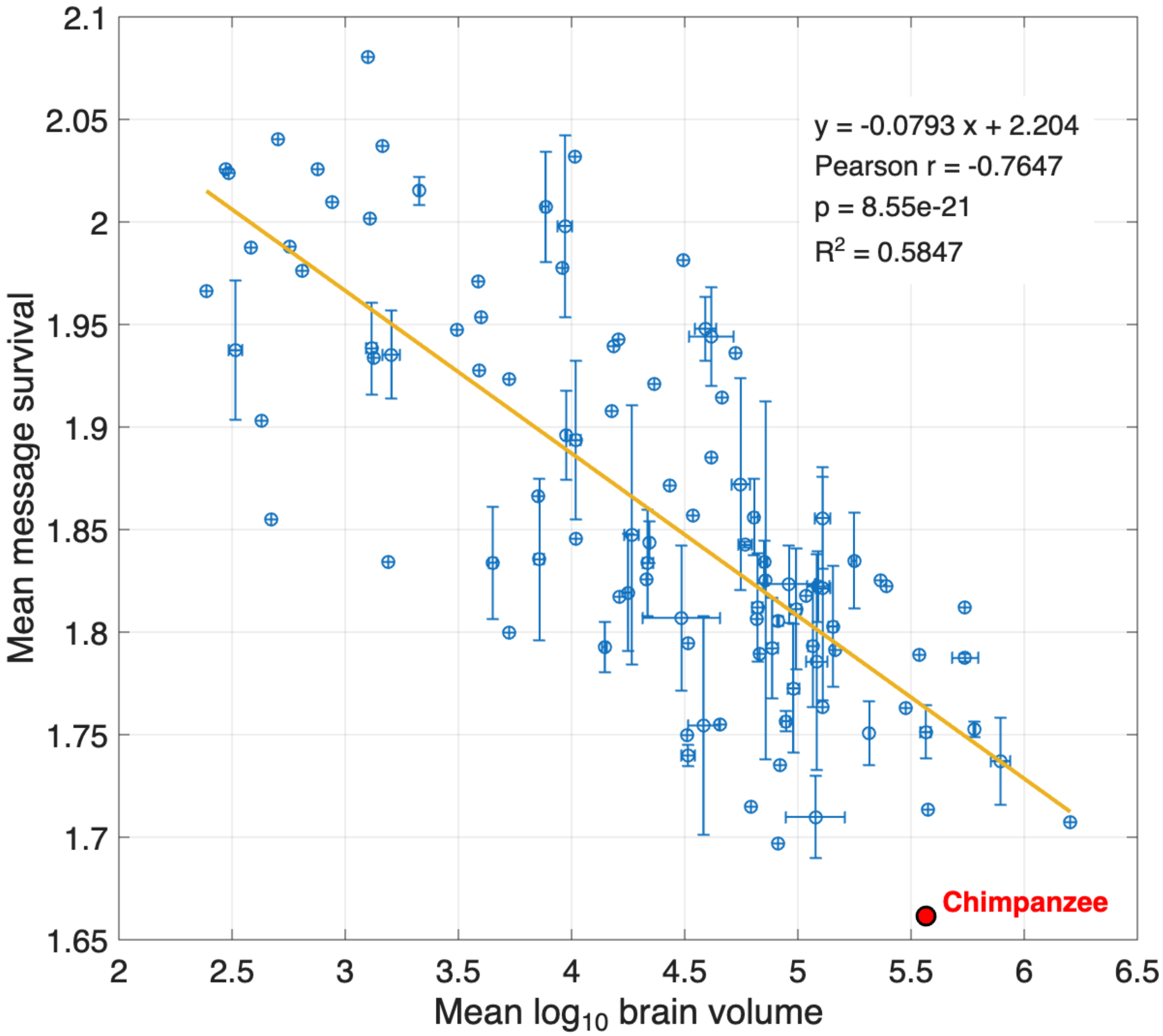

**Figure 4. Relationship between mean CSA message survival and mean $\log_{10}$ brain volume across 102 species (total of 185 mammalian connectomes analyzed; error bars indicate SEM).**

Under the CSA model, average message survival decreased strongly with increasing $\log_{10}$ brain volume (Pearson r = −0.7647, p = 8.6 × $10^{-21}$), accounting for over half of the variance across the 102 mammal species ($R^2$ = 0.585). See Figure 4. Chimpanzee, the species most closely related to humans in the dataset, had the lowest average message survival.

*Comparison with Alternative Measures*

We compared CSA message survival with a broad panel of established graph-theoretic measures and the colliding random-walk model in terms of their correlation with $\log_{10}$ brain volume.

Uncertainty in r and $R^2$ was estimated using a nonparametric species-level bootstrap with 10,000 replicates. In each replicate, the 102 species were resampled with replacement and the correlation recalculated. Ninety-five percent confidence intervals were defined by the 2.5th and 97.5th percentiles of the bootstrap distribution.

To compare CSA directly with each alternative measure, we used a paired species-level approach in which the same resampled species indices were applied simultaneously to brain volume, CSA message survival, and the alternative measure, thereby preserving the paired dependence among variables. For each bootstrap replicate, we computed the difference in absolute correlation coefficients, $\Delta|\mathrm{r}| = |\mathrm{r}_{\mathrm{CSA}}| - |\mathrm{r}_{\mathrm{alt}}|$, and the corresponding difference in $\mathrm{R}^2$, i.e., $\Delta\mathrm{R}^2 = (\mathrm{r}_{\mathrm{CSA}})^2 - (\mathrm{r}_{\mathrm{alt}})^2$.

To compare CSA directly with each alternative measure, we used a paired species-level approach in which the same resampled species indices were applied simultaneously to brain volume, CSA message survival, and the alternative measure, thereby preserving the paired dependence among variables. For each bootstrap replicate, we computed the difference in absolute correlation coefficients, $\Delta|r| = |r_{\mathrm{CSA}}| - |r_{\mathrm{alt}}|$, and the corresponding difference in $\mathrm{R}^2$, i.e., $\Delta\mathrm{R}^2 = (\mathrm{R}_{\mathrm{CSA}})^2 - (\mathrm{R}_{\mathrm{alt}})^2$. Two-sided bootstrap *p*-values were calculated as the proportion of bootstrap differences on either side of zero, and 95% confidence intervals were obtained from the percentile bootstrap distribution. Multiple comparisons across alternative measures were controlled using the Benjamini–Hochberg false discovery rate (FDR) procedure. An alternative measure was considered significantly weaker than CSA when the FDR-adjusted two-sided *q*-value was less than 0.05 and the corresponding 95% confidence interval for the paired difference excluded zero.

CSA message survival exhibited the largest observed absolute correlation with brain volume among all measures examined (Figure 5). Paired species-level bootstrap comparisons showed that CSA significantly outperformed the colliding random-walk model message survival, first-passage time, edge-based matching index, transitivity, spectral radius, total cover time, algebraic connectivity, closeness centrality, search information, degree assortativity, global efficiency, communicability, characteristic path length, Laplacian spectral radius, matching index (all pairs), and rich-club coefficient after false-discovery-rate correction ($q < 0.05$). Differences between CSA and clustering coefficient and path transitivity were not statistically significant after correction, indicating that these local neighborhood measures exhibited associations with brain volume statistically indistinguishable from that of CSA in the present dataset. Definitions of all measures and complete correlation statistics are provided in the Supplementary Materials.

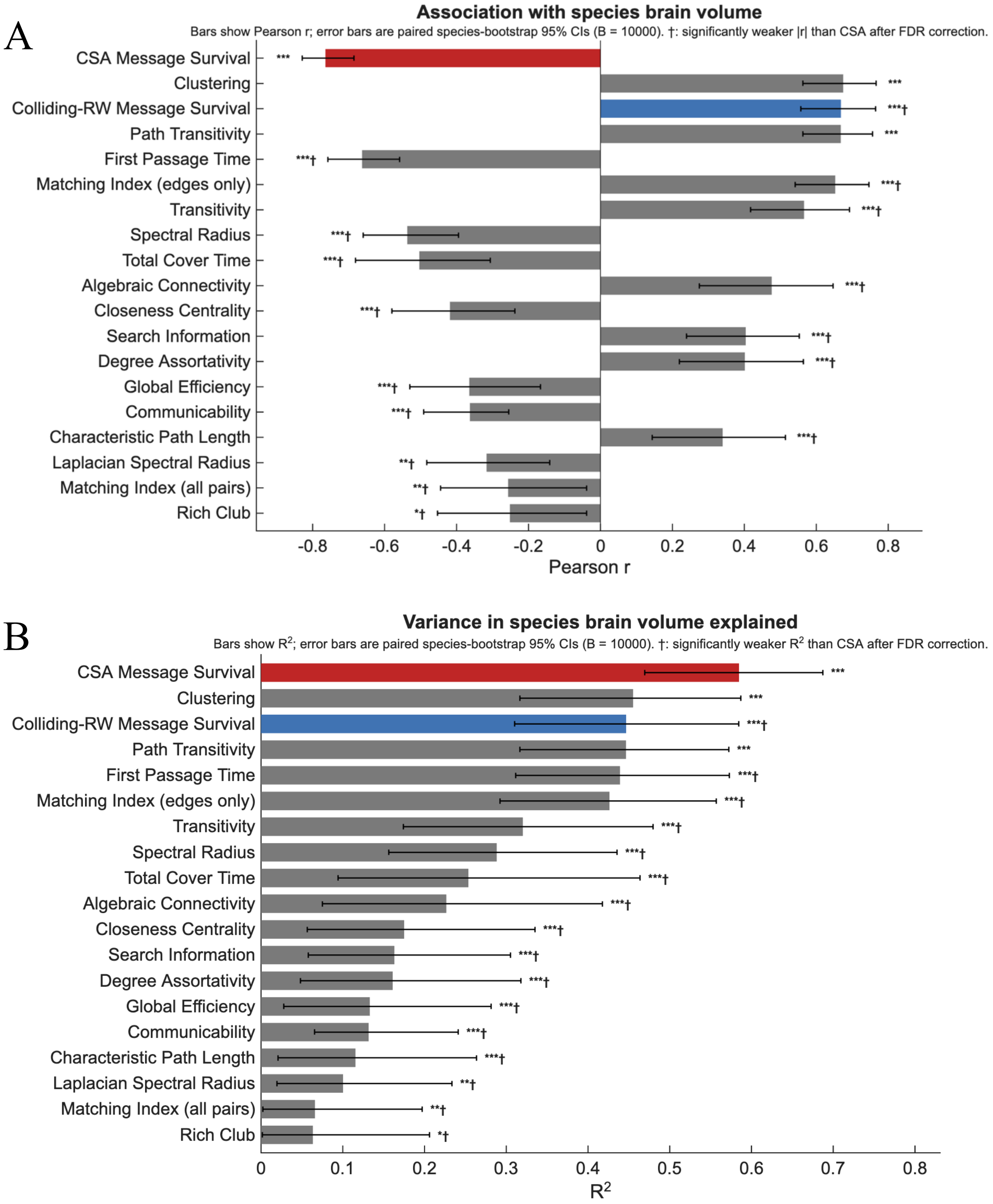

**Figure 5. A. Pearson correlation values for CSA message survival compared to other models and measures across 102 mammal species. B. Same data shown as $R^2$ values. Triple stars indicate p values less than 0.001, double stars indicate p values less than 0.01, single stars indicate p values less than 0.05. Definitions of alternative measures and a table of correlation values are given in the Supplementary Materials.**

Propagation Length, Spatial Reach, and Network Coverage

To assess whether our results depended on counting repeated node visits, we repeated the analysis using the longest loop-free injection-to-collision trajectory resulting from each injection. The underlying CSA dynamics were unchanged; only trajectories that never revisited a node qualified for this measurement of survival.

The preference for a lognormal-like survival distribution persisted when trajectories containing repeated node visits were excluded: the discretized lognormal model achieved the highest held-out log likelihood among the five candidate distributions in 98 of 102 species (96.1%), with the gamma distribution favored in the remaining four.

The negative association with $\log_{10}$ brain volume remained strong across the 102 species (Pearson $r = -0.7733$, $p = 1.6 \times 10^{-21}$, $R^2 = 0.598$; 95% bootstrap CI for $r = [-0.8385, -0.6926]$).

We then examined the spatial and tract-length properties of each injection's longest loop-free trajectories, defining the upper tail as maximum hop counts above the 90th percentile for each connectome. Averaging tied-longest trajectories within injections and weighting species equally, upper-tail paths more frequently visited nodes in the source-specific farthest Euclidean-distance decile (14.26% versus 5.43% for the remaining trajectories). Upper tail trajectories more frequently traversed at least one tract in the longest 1% in terms of length (1.38% versus 0.91% for the remaining trajectories), with both properties occurring on the same trajectory in 0.484% of cases (versus 0.187%). Thus, the lognormal-like long-hop tail included spatially long-range propagation without weights, tract lengths, or coordinates entering the propagation rules.

The probability that an injection reached at least one node in the source-specific farthest Euclidean-distance decile was positively associated with species mean message survival ($r = 0.4520$, $p = 1.9 \times 10^{-6}$) and negatively associated with $\log_{10}$ brain volume ($r = -0.6004$, $p = 2.6 \times 10^{-11}$). Thus, although total coverage and survival were not significantly correlated across species, survival was associated with the likelihood of reaching relatively distant destinations over physical space.

Together, these results support interpreting message survival as maximum propagation length per injection, rather than merely persistence through repeated circulation. The association with brain volume persists for non-repeating trajectories, and long trajectories include more long tracts and spatially distant visits.

We also quantified network coverage as the number of distinct nodes reached by any descendant of an injection, retaining looping trajectories and counting repeated arrivals only once. A node was considered reached when a descendant arrived, including arrivals that lead to collision; the initial injection location was excluded unless a descendant returned there subsequently. Injections reached an average of 46.57 nodes, or 23.28% of the network. Within connectomes, longer-lived injections generally reached more distinct nodes: the association was positive in all 185 connectomes, and the correlation after subtracting means from both variables was $r = 0.5216$ (95% CI = [0.5147, 0.5283]). However, mean coverage was not significantly associated with mean survival across species ($r = -0.1736$, $p = 0.081$), indicating that persistence and coverage capture related but distinct properties of propagation.

## Effects of Degree-Preserving Randomization

To determine whether the observed scaling relationship depends on higher-order network organization beyond degree sequence, we performed degree-preserving switch randomization (Maslov and Sneppen, 2002). We randomized each connectome in three partitions: the two hemispheres and the interhemispherical connections were randomized separately, with ~300,000 edge swaps per partition. Partitioning was done to equalize the level of interhemispheric connectivity with the corresponding empirical connectome; had we not done this, the switch randomization procedure would have increased the level of interhemispheric connectivity, and consequently changed the fundamental nature of the network (Fretter et al., 2012).

We generated five independent degree-preserving switch randomizations of each connectome and ran the CSA model using the same parameters as in the empirical analyses. We considered only randomizations that preserved full connectivity, and this retained 174 connectomes representing 96 species, each with five realizations. Because each connectome had five randomized realizations, we used Monte Carlo sampling to estimate the distribution of scaling relationships expected under the degree-preserving null model.

The empirical correlation of message survival with $\log_{10}$ brain volume for the 96 species was $r = -0.731$, while the randomized networks yielded a mean correlation of $r = -0.689$. Across the five randomized datasets, the relationship between message lifetime and brain volume was highly reproducible, with regression slopes ranging from −0.1002 to −0.0989 (mean = −0.0996, SD = 0.0006). The empirical slope for the corresponding set of 96 species (−0.0768) lay outside the range of all randomized realizations, indicating that the observed scaling relationship cannot be explained by degree sequence alone. Mean message lifetime was also significantly lower in the randomized than in the empirical networks (mean paired difference = −0.1085, $t(95) = -19.384$, $p = 8.81 \times 10^{-35}$).

We performed a Monte Carlo analysis using 10,000 iterations. In each iteration, one of the five randomized realizations was selected independently for each connectome. Species means were then recomputed using the same species assignments as in the empirical analysis, and the relationship between species mean message survival and $\log_{10}$ brain volume was recalculated. The Monte Carlo procedure generated the null distribution of the scaling relationship expected under a degree-preserving null model. The empirical correlation and slope both lay outside this null distribution (both $p = 1.0\times10^{-4}$), suggesting that degree sequence alone is insufficient to explain the observed relationship between message survival and brain volume. Higher-order features of empirical connectome organization therefore contribute measurably to the observed scaling.

## Structural Determinants of Message Survival

We next examined which structural properties are associated with message survival under CSA dynamics.

Hao et al. (2026) showed that neutral degree assortativity graphs maximize the trade-off between message creation and destruction engendered by different assortativity regimes, leading to maximal message lifetime. At high disassortativity, message spread is low, while message collisions are moderately high. At high assortativity, message spread is high, but collisions are very high. Complex networks with assortativity near zero strike a balance between spreading and collision, leading to maximal lifetime compared to graphs with the same degree sequence but different assortativity. Since all brain structural networks tested to date have assortativity near zero, we first examined assortativity as a structural determinant of message survival.

Although degree assortativity strongly modulates CSA dynamics under controlled graph manipulations (Hao et al., 2026), variation in degree assortativity across mammalian connectomes is relatively limited and explains only a modest fraction of the observed interspecific variation in message survival ($r = -0.3892$, $p = 5.3 \times 10^{-5}$). Degree assortativity shows a similarly modest correlation with brain volume ($r = 0.4012$, $p = 2.9 \times 10^{-5}$).

Unlike the random scale-free graphs and switch-randomized connectomes studied in Hao et al. (2026), the empirical mammal connectome dataset studied here was not equalized in terms of degree sequence across species (though the number of nodes and edges was constant, and the randomized connectomes studied above have the same degree sequence as the corresponding empirical ones). Therefore, we expected that variability in degree distributions would influence message survival in ways that differed from the mechanistic model proposed by Hao et al. (2026). Because degree sequence was allowed to vary across species in the present dataset, we asked whether simple properties of the degree distribution predicted message survival.

The mean degree of hubs (nodes above the $90^{th}$ percentile in each connectome, averaged within species) was strongly associated with message survival ($r = 0.7179$, $p = 2.0 \times 10^{-17}$). However, mean hub degree was substantially less strongly associated with brain volume ($r = -0.5188$, $p = 2.3 \times 10^{-8}$) than message survival itself ($r = -0.7647$). A paired bootstrap comparison confirmed that the difference in correlation magnitude was significant ($\Delta|r| = 0.2458$, 95% CI = [0.1310, 0.3772], one-sided $p < 10^{-4}$). Correspondingly, CSA message lifetime explained 58.5% of the variance in brain volume compared with 26.9% for mean hub degree ($\Delta R^2 = 0.3155$, 95% CI = [0.1795, 0.4418], one-sided $p < 10^{-4}$). Thus, a structural feature that strongly predicts the CSA dynamics nevertheless predicts brain volume substantially less well than the dynamics themselves.

Node-Level Organization of Collisions

To relate global message survival to local network organization, we examined the distribution of collisions across nodes, defining collision frequency as the fraction of timesteps in which a node received two or more message copies. Full methods and results are given in the Supplementary Materials and summarized here. Within-connectome correlations, summarized across species, showed that collision frequency was strongly positively associated with degree ($r = 0.8394$) and negatively associated with local clustering ($r = -0.5660$). Hubs, defined as nodes strictly above the 90th degree percentile within each connectome, constituted approximately 9.5% of nodes (ties at the threshold were excluded), but accounted for 19.2% of all deleted copies. Thus, frequent collisions and disproportionate message deletion were associated with high degree, whereas high local clustering was associated with less frequent collisions.

The participation coefficient, which measures how broadly and evenly a node's connections are distributed among structurally defined modules, was positively associated with total collision frequency, but this association was much weaker after linear adjustment for degree, and reversed after rank-based adjustment for degree. In contrast, participation was strongly associated with collisions involving simultaneous arrivals from immediate senders in different modules ($r = 0.8655$), including after degree adjustment. Thus, degree primarily characterized collision frequency, whereas intermodular connectivity characterized cross-module convergence. Restricting these events to collisions containing at least one pair of arrivals that differed in both immediate-sender module and injection lineage left the association essentially unchanged ($r = 0.8628$), including after degree adjustment, indicating that connector nodes preferentially hosted collisions of messages from distinct injection lineages across modules beyond the association accounted for by degree.

Because the connectomes have neither functional assignments nor anatomical correspondence across species/connectomes, these analyses describe collision likelihood in terms of topological features but cannot determine their location in cortical hierarchies or assess whether particular anatomical collision sites are conserved across species.

## DISCUSSION

### *Principal Findings*

We investigated how interactions among simultaneously propagating messages influence communication on unweighted, undirected, non-spatially embedded mammalian connectomes.

Three principal findings emerged. First, CSA dynamics generate robust lognormal message survival distributions across species despite relying only on binary network topology (including in the loop-free case). Second, message survival accounts for over half the variance in mammalian brain volume, while structural features—including those that influence message survival under the CSA dynamical model—are not as strongly correlated with brain volume compared to the dynamical measure itself. Third, degree-preserving randomization weakens this relationship, suggesting that the CSA model's emergent dynamics, in concert with the higher-order network organization, contributes to the observed effects.

Message interactions on a network could in principle generate message survival of a great variety of distributions, given the right structural and dynamical parameters. The EDR does not itself provide a prediction of survival distributions since it does not consider message interactions or polysynaptic message travel. However, it is reasonable to expect that the net effect of the EDR on each edge, and in turn on network topology, should give rise to EDR-like distributions of message survival.

Thus, having many short connections available for each area may in part serve to allow redundant messages to travel more easily across the network, given the presence of message interactions. The results of the first experiment show that the CSA model produces dynamic behavior in line with what one would expect based on pervasive lognormal-like structure and dynamics in the brain when one includes message interactions.

Under CSA, larger brain volume was associated with shorter maximum propagation lengths and a lower probability of reaching relatively distant destinations. Colliding-RW survival showed the opposite association with brain volume, demonstrating that the scaling of persistence depends on the propagation dynamics when messages can interact. In comparison to past findings on static network measures, our results suggest that the localization of communication in bigger brains is found under plausible message interaction dynamics, implying that it is not just increasing spatial separation that keeps interacting messages local but also graph topology. The effects of network structure on the propagation of interacting messages are emergent and self-limiting, suggesting that they could ultimately relate to metabolic budgets in brain development and evolution.

CSA-like strategies may be useful not only in mesoscale white matter networks, but also brains in orders outside Mammalia (e.g., fly: Dorkenwald et al., 2024, which we are currently investigating), as well as micro-scale neuronal networks, which show similar topology (Guidolin et al. 2016; Grosu et al., 2023).

*Stochastic Communication*

The CSA model represents a stochastic communication framework in which routing emerges from local interactions rather than centralized control or pre-specified paths. We note that many other classes of system-wide models of signal flow in brains take a dynamic approach, including oscillator models (e.g., Benitez-Stulz et al., 2024), traveling wave models (e.g., Keller et al., 2024), neural mass models (e.g., Breakspear, 2017), and excitable systems (e.g., Kaiser et al., 2007; Kaiser and Hilgetag, 2010; Messé et al., 2015). However, these models have not generally addressed the communicatory goals of the system, and they have mostly ignored the question of how coherent messages could survive multiple synaptic traversals if messages can interact with each other.

Our model has several advantageous properties, not least that it requires no central controller or path-finding mechanism to achieve stable, emergent, adaptive communication. Massive message redundancy obviates the need for addressing, error correction, and node buffers, which are required by queueing models (see Supplementary Materials) but not proven to exist in mammalian brains. The seemingly profligate strategy of copying each surviving message to all neighbors serves to offset deletions at collision. In past tests, we have found that less drastic collision rules (e.g., allowing one message to propagate onward from a collision) lead to rapid overload of the system (Hao and Graham, 2020).

Stochasticity in the CSA model stems from the injection rule, which is random and unbiased. The fact that we do not need to bias inputs to see robust relationships with biological properties suggests that intrinsic activity in the brain may be more meaningful and fundamental to brain function than is sometimes believed (see e.g., Buzsaki, 2019). However, biased implementations are worth exploring as well.

*Biological Plausibility*

Although the CSA model is not a biophysical model, discrete messaging in the brain is biologically plausible given single-unit electrophysiological evidence for "message packets" being passed from neuron to neuron over the course of tens or hundreds of milliseconds. "Messages" of this kind have been observed in rodent hippocampus (Nádasdy et al., 1999; Grosmark and Buzsaki, 2016) and in auditory (Hahnloser et al., 2002; Hromadka et al., 2008; Luczak et al., 2013; 2015), somatosensory (Foffani et al., 2008), and olfactory cortex (Junek et al., 2010). Although definitions of what constitutes a message in these studies vary, there is enough suggestive evidence for their existence to motivate study of abstract models of discrete message exchange on brain networks.

We suggest that a core network communication task that brains perform is the creation and—crucially—the deletion of polysynaptic messages. On the network of white matter tracts, which is almost entirely excitatory, we presume that internal node dynamics within regions act to cancel out signals that arrive simultaneously from different incoming edges. Deletion could occur via classical inhibition and any of a variety of nonlinear mechanisms (Gidon et al., 2020; 2022; Steriade and Paré, 2007; Gollisch and Meister, 2010; Oz et al., 2022).

*Coincidence Detection*

We hypothesize that message deletion serves functionally as the sensing of coincidences, i.e., detecting events that impinge on that area together. Successful communication constitutes nodes being aware of instances when messages arrive on incoming edges at the same time, which arose at two or more different locations on the network and, potentially, at different times. Once delivered, this "multi-message" should not propagate further. This goal is distinct if not independent from the goal of transferring specific functional or content information from one area to another, such as from one level of visual system analysis (e.g., lateral inhibition in the retina) to the next (e.g., edge detection in V1).

Instead of passing along even more excitation due to the coincident incoming excitations, the outcome of a collision under the CSA model is to delete all coincident excitations. This is a major difference compared to proposed coincidence detector circuits, and indeed most computational models of the brain, including most dynamical systems models of the brain. If excitations are treated as messages, the successful arrival of coincident messages in such models is always a trigger for more messages to be sent. Yet previous models have not dealt directly with the network communication needs of the nervous system, specifically congestion management and message delivery, which we argue require systematic message deletion.

In this view, network architecture and signal timing determine *how* and *where* information is combined, while the internal machinery of a brain area serves to determine what the coincidence is *of*. This interpretation is bolstered by the finding that connector nodes preferentially receive and terminate simultaneous inputs arriving from different modules. While this result concerns convergence under CSA dynamics, not demonstrated biological coincidence detection or complementary information content, we suggest that this general framework could apply to, for example, auditory and visual cues that coincide in time, fragments of memory traces with related content that are brought together during recall, or descending motor commands that generate muscle synergies. Each batch of coincident signals may require different computational processing depending on the signals and the brain area involved. But this shouldn't make us lose sight of the network communication goal, which is to detect coincidences.

Thus, one reason why evolution has maintained long-range connections, which become increasingly costly for larger brains (Zhang and Sejnowski, 2000; Laughlin and Sejnowski, 2003; Bullmore and Sporns, 2012; Karl et al., 2024), may be to ensure combinatoric diversity of simultaneous message arrivals at nodes in the presence of potentially deleterious message interactions. Under this assumption, the nature of the brain's topology can be seen to impose a constraint on how far a given message can propagate in a mammalian brain before it meets complementary messages.

*How Far Should a Message Travel On the Connectome?*

A basic design specification for any communication network is how long messages typically survive on the network. Some systems, such as 19$^{th}$ c. semaphore in France and 15$^{th}$ c. Yam postal systems in Eurasia, largely generated long-lived messages, which usually traveled from stations on the periphery of the network to a single central station, and vice versa (Gleick, 2011). Other communication systems such as human speech are strictly local and very short-lived: one's voice can only be heard transiently and by those in the immediate vicinity. The internet permits a wide range of message propagation distances with negligible increases in delays even at long distances.

Though neuronal excitations are much slower than the internet's signals, the brain likely also requires that message passing can succeed over a range of network distances and times in order to achieve multisystem integration (Danilova and Mollon, 2003; Mollon and Danilova, 2019; Mollon et al., 2022; Graham, 2014; 2021; 2023) and temporal coherence (Zeki and Bartels, 1998; Zeki, 2020). The maintenance of long-distance connections, however few, in the largest brains, despite the steeply increasing cost of building such connections, suggests that they are an indispensable part of brain design. We therefore argue that understanding message survival on brain networks is a key step in elucidating the nature of evolved communication strategies over mammal brain networks.

The present results provide intuition about how far a putative discrete message "should" travel on brain networks, even if we cannot say what constitutes a message. Thus, our result suggests a testable prediction. If a quantity defining a message can be specified, one would look across neuronal populations for all copies of a given message and ask how many hops the longest-lived copies take. We predict that such measures will produce distributions that resemble the distributions of message survival under the CSA model. In addition, we predict that there is a natural cut-off in message survival for any brain network—some number of hops or some period of survival that messages do not usually exceed.

*Limitations*

Important and well-documented limitations to several aspects of the connectome data acquisition and processing pipeline are described by Puxeddu et al. (2024). For example, there are challenges to acquiring tractography on a common basis in brains that span more than three orders of magnitude in size and were acquired under differences in field strengths and other imaging parameters. This causes difficulty both for the definition of nodes ("functional" brain areas, e.g., left V1) and of edges/tracts. Thresholding the network (i.e., iteratively removing edges with the smallest weights to achieve a fully connected graph of a fixed edge count for each brain) imposes additional limitations as discussed in Puxeddu et al., (2024) and Faskowitz et al. (2022); see also Van Wijk et al. (2010). However, the findings by Faskowitz et al. (2022) and Suárez et al. (2022) that phylogenetic relationships among mammals can be reliably reconstructed from networks in the MaMI dataset adds significant support for the utility and anatomical accuracy of these networks.

The CSA model is intentionally abstract and should not be interpreted as a biophysical model of neuronal signaling. Nevertheless, results of simulations using our model show novel, intriguing, and biologically relevant patterns of network communication on the mammalian connectome. Our findings are consistent with known network regularities in the brain, and our model makes predictions for features of brain activity.

Numerical simulations of the CSA model generated patterns indicative of brains, but without explicit mechanistic designs for those patterns. This makes them useful, especially given the difficulty of analytical solutions to the nonlinear dynamics we studied. In other domains, success in simulation or in technological practice has likewise served as useful evidence, especially when theoretical underpinnings are lacking or difficult to determine (see e.g., Emmert-Streib et al., 2024). Consider the internet. Though each communication linkage on the internet can be analyzed in information theoretic terms, the system's robustness and efficiency are proven by demonstration rather than by analysis (Pastor-Satorras and Vespignani, 2004; El Gamal and Kim, 2011). The same applies in parallel for artificial neural network models: insights are gained through numerical simulation and functional performance, rather than on formal analysis of internal mechanisms, which remain obscure despite much study. Nevertheless, analytical solutions are desirable.

The finding that larger brains show shorter message survival times is in line with studies of static, linear network statistics in the brain. Work by Puxeddu et al. (2024) found that "as the brain volume increases, modules become more spatially compact and dense, comprising more costly connections." The fact that this behavior is implied by measures of structural features does not necessarily imply that dynamical, nonlinear behavior follows directly from network structure. For collision models to match brains, we may require spreading activation since the colliding-RW model generated fundamentally different behavior. However, there exists an unknown space of routing schemes with biological plausibility that we have not explored; we cannot exclude the possibility that different spreading and/or collision models produce similar results.

Indeed, the space of possible routing strategies deserves systematic study. Models that employ elements of spreading, return messages (acks), resending, buffers, and/or collisions are possible (Hao and Graham, 2018; Graham, 2023), and more sophisticated models will need to consider regional variations in message passing needs and capabilities. Nevertheless, understanding simple, uniform, stochastic models will help guide us toward more complex and realistic models. A new basic science of routing may be needed to understand why brain networks—as well as other biological and engineered networks—are such highly effective communication systems (see Ji et al., 2023; Newman, 2023; Paz, 2022).

## CONCLUSION

The present work suggests that communication in mammalian brain networks cannot be understood solely from static topology considerations. Interactions among propagating messages introduce emergent dynamics that shape communication over the connectome and scale systematically with brain organization across species. Whether or not the brain implements CSA dynamics specifically, our results indicate that message interactions deserve consideration as a fundamental component of future theories of large-scale neural communication.

## ACKNOWLEDGMENTS

We are grateful to the creators of the Brain Connectivity Toolbox (Rubinov and Sporns, 2010), whose Matlab functions were used. We thank Walden Marshall, Randy Hong, Donald Spector, Chris Fietkiewicz, Joe Rusinko, Bruno Olshausen, Barbara Finlay, David Field, Colin Conwell, Frank Tong, Gorka Zamora-López, and Peter Latham for helpful discussions. We dedicate this work to the memory of Erik Bollt and Barbara Finlay.

## COMPETING INTERESTS

The authors have declared that no competing interests exist.

# SUPPLEMENTARY MATERIALS

## DISCUSSION OF QUEUEING MODELS

The study of message interactions in the brain takes inspiration in part from engineered networks like the Internet, which is extremely efficient at dealing with congestion (Kleinrock, 1976; Mollon and Danilova, 2018; Graham et al., 2021). And indeed, the first models to test the effects of message interactions on discrete messages were *queueing models*, which treat nodes (brain areas) essentially as routers with memory buffers (Misic et al., 2014a,b).

Queueing models assume that nodes store messages in a local memory buffer, and that messages are passed to pre-specified target nodes according to a fixed rule applied to all nodes, such as first-in, first-out (FIFO) or last-in, first-out (LIFO) queueing. In these models, messages are discrete entities, while system dynamics are based on a Poisson event process that is biased by edge weight. Simulations of these models on a directed macaque monkey connectome have suggested that some brain areas such as the hippocampus handle substantially more message traffic than would be predicted based on local connection strength alone, presumably because of longer-range interactions of structural and dynamical factors (Misic et al., 2014a,b; see also Graham, 2014).

Follow-up work using simulations on the macaque and mouse connectomes by Fukushima and Leibnitz (2024) also found benefits to congestion management in the sense of promoting efficient message-passing between a given source and its target. Benefits were found not just in utilizing memory buffers for discrete messages, but also in dividing messages into multiple packets. Note that here the term packets is consistent with Internet packets, in which a “message” consists of a collection of equal-sized pre-addressed packets, whereas electrophysiological studies that invoke this term tend to use it to mean discretized polysynaptic signals (e.g., Luczak et al., 2015; see Discussion).

Perhaps the most intriguing finding in the work of Fukushima and Leibniz (2024) is that packetization is about as beneficial for efficiency (defined as whole-message transit time) when whole messages were all the same size as when whole message sizes varied. This suggests that a basic routing principle—packetization—underlying the Internet’s efficiency could also contribute to efficient principles of brain network communication. As with the Internet, whole messages in the brain may come in different sizes.

These models have been important in identifying how congestion and its management can influence message flows. However, no plausible addressing system has been proposed. And although buffers are biologically plausible as models of short-term memory in the brain (Goldman-Rakic, 1996; Funahashi, 2015; Li and van Rossum, 2020), potentially through local recurrent loops, buffers as such have not been directly observed in mammalian brains. Moreover, queueing models perform best when buffers are unlimited in size, rather than small or even finite. The benefits of packetization—chiefly, fast and reliable delivery at targets—are largely eliminated if packets are lost when buffers are full, as both Misic et al. (2014a,b) and Fukushima and Leibnitz (2024) found. This problem is compounded in the Fukushima and Leibnitz (2024) model by the fact that losing just one packet potentially corrupts the entire message. More generally, it is not clear that brain network communication has source-target communication (i.e., “two-port” communication that is the object of study for Shannon’s information theory) as its goal.

## ALTERNATIVE NETWORK MEASURES AND MODELS

Let $G = (V, E)$ be an undirected binary graph with N nodes; $A$ is the adjacency matrix; $D$ is the degree matrix; The Laplacian $L = D - A$; $k_i$ is node degree; and $d$(i,j) is shortest-path distance.

**Clustering**
The local clustering coefficient quantifies the fraction of pairs of neighbors of node i that are themselves connected; the graph value is the mean across nodes.

$$C_i = \frac{2T_i}{k_i(k_i-1)}, \qquad C = \frac{1}{N}\sum_{i=1}^{N} C_i$$

**Path Transitivity**
Path transitivity measures the density of matching relationships among node pairs lying on a shortest path, averaged across paths.

$$PT(i,j) = \frac{1}{|\mathcal{P}_{ij}|}\sum_{(u,v)\in\mathcal{P}_{ij}} M_{uv}, \qquad PT = \frac{1}{N(N-1)}\sum_{i\neq j} PT(i,j)$$

**First Passage Time**
Mean first-passage time is the expected number of steps for an unbiased random walker starting at node i to visit node j for the first time.

$$m_{ij} = \frac{Z_{jj}-Z_{ij}}{\pi_j}, \qquad Z = (I - P + \mathbf{1}\pi^T)^{-1}, \qquad MFPT = \frac{1}{N(N-1)}\sum_{i\neq j} m_{ij}$$

**Matching Index (edges only)**
The matching index quantifies neighborhood overlap; here it is averaged only over connected node pairs.

$$M_{\text{edge}} = \frac{1}{|E|}\sum_{(i,j)\in E}\frac{|N(i)\cap N(j)|}{|N(i)\cup N(j)|}$$

**Transitivity**
Transitivity is the fraction of connected triples that are closed into triangles.

$$T = \frac{3N_{\triangle}}{N_{\text{triples}}}$$

**Spectral Radius**
The spectral radius is the largest absolute eigenvalue of the adjacency matrix.

$$\rho(A) = \max_i |\lambda_i(A)|$$

**Total Cover Time**

Cover time is the number of random-walk steps required to visit every node; the graph value is averaged over starting nodes.

$$C = \frac{1}{N} \sum_{i=1}^{N} C_i$$

**Algebraic Connectivity**

Algebraic connectivity is the second-smallest eigenvalue of the combinatorial Laplacian.

$$\lambda_2(L), \qquad L = D - A$$

**Closeness Centrality**

Closeness centrality is the inverse of the mean shortest-path distance from a node to all other nodes; the graph value is averaged across nodes.

$$CC_i = \frac{N-1}{\sum_{j \neq i} d(i,j)}, \qquad CC = \frac{1}{N} \sum_{i=1}^{N} CC_i$$

**Search Information**

Search information is the negative base-2 logarithm of the probability that an unbiased random walker follows a shortest path exactly.

$$SI(i,j) = -\log_2 \left[ \prod_{v \in \pi_{ij}^-} \frac{1}{k_v} \right], \qquad SI = \frac{1}{N(N-1)} \sum_{i \neq j} SI(i,j)$$

**Degree Assortativity**

Degree assortativity is the Pearson correlation between the degrees at the two ends of an edge.

$$r = \frac{\sum_e (x_e - \bar{x})(y_e - \bar{y})}{\sqrt{\sum_e (x_e - \bar{x})^2 \sum_e (y_e - \bar{y})^2}}$$

**Global Efficiency**

Global efficiency is the mean inverse shortest-path distance over all ordered node pairs.

$$E_{\text{glob}} = \frac{1}{N(N-1)} \sum_{i \neq j} \frac{1}{d(i,j)}$$

**Communicability**

Communicability weights all walks between two nodes by the inverse factorial of walk length; the graph value is the mean off-diagonal entry.

$$G = e^A, \qquad \overline{G} = \frac{1}{N(N-1)} \sum_{i \neq j} G_{ij}$$

**Characteristic Path Length**

Characteristic path length is the mean shortest-path distance over all ordered node pairs.

$$L = \frac{1}{N(N-1)} \sum_{i \neq j} d(i,j)$$

**Laplacian Spectral Radius**

The Laplacian spectral radius is the largest eigenvalue of the combinatorial Laplacian.

$$\rho(L) = \lambda_{\max}(L)$$

**Matching Index (all pairs)**

The all-pairs matching index is the mean neighborhood overlap across all unordered node pairs.

$$M_{\mathrm{all}} = \frac{2}{N(N-1)} \sum_{i<j} \frac{|N(i) \cap N(j)|}{|N(i) \cup N(j)|}$$

**Rich Club**

The rich-club coefficient at degree threshold k is the density of connections among nodes with degree greater than k.

$$\phi(k) = \frac{2E_{>k}}{N_{>k}(N_{>k}-1)}$$

**TABLE OF PEARSON CORRELATIONS WITH BRAIN VOLUME**

| Measure/Model | Pearson r | p |
|---|---|---|
| CSA Message Survival | −0.7647 | 8.60E−21 |
| Clustering | 0.6747 | 7.54E−15 |
| Colliding Random Walk | 0.6684 | 1.65E−14 |
| Path Transitivity | 0.6682 | 1.69E−14 |
| First Passage Time | −0.6627 | 3.28E−14 |
| Matching Index (edges only) | 0.6529 | 1.04E−13 |
| Transitivity | 0.5660 | 5.68E−10 |
| Spectral Radius | −0.5370 | 5.95E−09 |
| Total Cover Time | −0.5037 | 6.82E−08 |
| Algebraic Connectivity | 0.4762 | 4.23E−07 |
| Closeness Centrality | −0.4184 | 1.21E−05 |
| Search Information | 0.4040 | 2.55E−05 |
| Degree Assortativity | 0.4012 | 2.92E−05 |
| Global Efficiency | −0.3648 | 1.63E−04 |
| Communicability | −0.3628 | 1.79E−04 |
| Characteristic Path Length | 0.3398 | 4.75E−04 |
| Laplacian Spectral Radius | −0.3167 | 1.20E−03 |
| Matching Index (all pairs) | −0.2569 | 9.20E−03 |
| Rich Club | −0.2519 | 1.07E−02 |

**TOPOLOGICAL LOCALIZATION OF MESSAGE COLLISIONS**

To connect global message survival with local network organization, we examined whether collision events preferentially occurred at high-degree hubs, within densely interconnected neighborhoods, or at nodes connecting different modules. We recorded nodewise collisions across all 185 connectomes using the same CSA parameters as in the preceding analyses: 13 injections per timestep, 2,000 timesteps, and 10 runs per connectome, with looping trajectories retained. The injection rate corresponded to 6.5% of the total number of nodes, with injection sites randomly selected from unoccupied nodes. Collision frequency was defined as the fraction of timesteps in which a node received two or more message copies. Each such node–timestep was counted as one event, irrespective of how many copies collided; the number of copies annihilated was recorded separately.

For each connectome, we correlated collision frequency across its 200 nodes with degree, local clustering coefficient, and participation coefficient. Participation coefficient quantifies how broadly and evenly a node's connections are distributed among modules (Guimerà and Amaral, 2005). Modules were identified by repeating Louvain community detection 25 times per connectome with different random seeds and retaining the partition with the highest modularity at resolution $\gamma = 1$ (Blondel et al., 2008). Correlations were Fisher-z transformed, averaged across connectomes within species, averaged equally across the 102 species, and then back-transformed. Ninety-five percent confidence intervals were obtained from 10,000 species-bootstrap resamples. All correlation coefficients reported below summarize within-connectome associations, rather than treating nodes pooled across brains as independent observations.

Collision frequency was strongly positively associated with node degree ($r = 0.8394$, 95% CI = [0.8331, 0.8455]; Spearman $\rho = 0.9717$), with positive associations in all 185 connectomes. The especially strong rank correlation indicated that degree closely ordered nodes by collision frequency, although collision frequency saturated at high degree. Hubs, defined as nodes strictly above the graph-specific 90th degree percentile, experienced collisions on 98.8% of timesteps, compared with 79.1% at other nodes. Thus, collisions were widespread but occurred most consistently at hubs.

The concentration of message deletion at hubs was more pronounced when counting annihilated copies rather than collision events. Degree was strongly associated with the number of copies deleted per timestep ($r = 0.9690$, 95% CI = [0.9677, 0.9702]). Hubs constituted approximately 9.5% of nodes (ties at the degree threshold were excluded) but accounted for 19.2% of all deleted copies. An average of 8.89 copies was deleted per hub per timestep, compared with 3.97 at other nodes. These results distinguish collision frequency from message loss.

In contrast, local clustering was negatively associated with collision frequency ($r = -0.5660$, 95% CI = [−0.6073, −0.5217]), with negative associations in 179 of 185 connectomes. Nodes above the 90th clustering percentile experienced collisions on 57.8% of timesteps, compared with 83.5% at remaining nodes. Thus, high local clustering did not identify preferential collision locations.

Participation coefficient was positively associated with total collision frequency ($r = 0.5136$, 95% CI = [0.4986, 0.5286]), with positive associations in all 185 connectomes. However, this association was strongly dependent on degree: linear adjustment (regressing each node's collision frequency and participation coefficient with degree in each connectome and examining the correlation of residuals) yielded a small positive partial correlation ($r = 0.1040$), while performing the same procedure over ranks gave a small negative correlation ($\rho = -0.1629$). The unadjusted relationship therefore does not establish that participation contributes positively to total collision frequency beyond degree.

A different pattern emerged for cross-module collisions, defined as simultaneous arrivals from immediate sending nodes belonging to at least two modules. Participation was strongly associated with the frequency of these events ($r = 0.8655$, 95% CI = [0.8576, 0.8729]), even after linear and rank-based degree adjustment ($r = 0.8616$ and $\rho = 0.8211$, respectively), and were positive in every connectome. Thus, degree primarily characterized how frequently nodes experienced collisions, whereas participation strongly characterized the convergence of arrivals from different modules.

To distinguish convergence of separately-injected messages from convergence of copies sharing a common injection, we restricted cross-module events to those containing at least one pair of arrivals that differed in both immediate-sender module and injection lineage. This restriction left the correlation with participation essentially unchanged ($r = 0.8628$, 95% CI = [0.8549, 0.8704]; linear degree adjusted $r = 0.8601$; degree rank adjusted $\rho = 0.8232$), again with positive associations in all 185 connectomes.

Collectively, these analyses distinguish frequent collisions and disproportionate message deletion at high-degree nodes from cross-module convergence of distinct injection lineages at high-participation nodes. Such convergence is consistent with the proposed interpretation of collisions as coincidence events shaped by network architecture, but does not yet establish specialized information integration. Moreover, the modules were defined structurally rather than by anatomical homology or functional specialization.